\documentstyle[12pt]{article}
\makeatletter\newdimen\normalarrayskip            % skip between lines
\newdimen\minarrayskip                 % minimal skip between lines
\normalarrayskip\baselineskip
\minarrayskip\jot
\newif\ifold             \oldtrue            
\def\arraymode{\ifold\relax\else\displaystyle\fi}
\def\eqnumphantom{\phantom{(\theequation)}}
\def\@arrayskip{\ifold\baselineskip\z@\lineskip\z@
     \else
     \baselineskip\minarrayskip\lineskip2\minarrayskip\fi}
\def\@arrayclassz{\ifcase \@lastchclass \@acolampacol \or
\@ampacol \or \or \or \@addamp \or
   \@acolampacol \or \@firstampfalse \@acol \fi
\edef\@preamble{\@preamble
  \ifcase \@chnum
     \hfil$\relax\arraymode\@sharp$\hfil
     \or $\relax\arraymode\@sharp$\hfil
     \or \hfil$\relax\arraymode\@sharp$\fi}}
\def\@array[#1]#2{\setbox\@arstrutbox=\hbox{\vrule
     height\arraystretch \ht\strutbox
     depth\arraystretch \dp\strutbox
     width\z@}\@mkpream{#2}\edef\@preamble{\halign \noexpand\@halignto
\bgroup \tabskip\z@ \@arstrut \@preamble \tabskip\z@ \cr}%
\let\@startpbox\@@startpbox \let\@endpbox\@@endpbox
  \if #1t\vtop \else \if#1b\vbox \else \vcenter \fi\fi
  \bgroup \let\par\relax
  \let\@sharp##\let\protect\relax
  \@arrayskip\@preamble}
\def\eqnarray{\stepcounter{equation}%
              \let\@currentlabel=\theequation
              \global\@eqnswtrue
              \global\@eqcnt\z@
              \tabskip\@centering
              \let\\=\@eqncr
              $$%
 \halign to \displaywidth\bgroup
    \eqnumphantom\@eqnsel\hskip\@centering
    $\displaystyle \tabskip\z@ {##}$%
    &\global\@eqcnt\@ne \hskip 2\arraycolsep
         \hfil$\arraymode{##}$\hfil
    &\global\@eqcnt\tw@ \hskip 2\arraycolsep
         $\displaystyle\tabskip\z@{##}$\hfil
         \tabskip\@centering
    &{##}\tabskip\z@\cr}
\makeatother

\def\beq{\begin{equation}}
\def\eeq{\end{equation}}
\def\bea{\begin{eqnarray}}
\def\eea{\end{eqnarray}}

\begin{document}
\begin{titlepage}
\begin{center}
\hfill VU-04/93\\
\hfill UUITP  $34$/1993\\
\hfill hep-th/9404160\\
%\begin{flushright}{17 April 1993}\end{flushright}
\vspace{0.7in}{\Large\bf 3nj-symbols and D-dimensional \\
quantum gravity}\\[.4in] {\large P.Demkin}\footnote{On leave
from Department of Physics, Vilnius University, Saul\.{e}tekio al.9,
2054, Vilnius, Lithuania}\\
\bigskip {\it
Institute of Theoretical Physics \\ Uppsala University\\ Box 803,
S-75108\\ Uppsala, Sweden\\ paul@rhea.teorfys.uu.se}

\end{center}
\bigskip \bigskip

\begin{abstract}
The model which generalizes Ponzano and Regge $3D$ and
Carfora-Martellini-Marzuoli
$4D$ euclidean quantum gravity is considered. The euclidean Einstein-Regge
action for a
$D$-simplex is given in the semiclassical limit by a gaussian integral of a
suitable
$3nj$-symbol.
\end{abstract}

\end{titlepage}
\newpage
%\section{Introduction}

Some time ago G.Ponzano and T.Regge discovered a connection between
 the expansion of the $6j$-symbol and the partition function for $3D$
 euclidean quantum gravity \cite{PR}. These ideas were developed by
 M.Carfora, M.Martellini and A.Marzuoli for the $4D$ euclidean quantum gravity
 \cite{CMM}. In this paper the general correlation between $3nj$-symbols and
$D$
 euclidean quantum gravity is shown.

The main aspect here is the correlation between geometrical properties of the
 $D$-simplex and the corresponding $3nj$-symbol. The $D$-simplex in the
$D$-dimentional space has $D+1$ vertices and $C^{2}_{D+1}=\frac {1}{2}D(D+1)$
edges.
No less than
 $D-3$ edges belong to $D-3$-dimensional subspace and no more than
 $\frac {1}{2}D(D-1)+3$ edges belong to 3-dimensional subspace $R^3$. To every
edge
belonging to $R^3$ we relate some quantity of the angular momenta $j_i$ in the
 $3nj$-symbol. Therefore, $3nj$-symbol must contain no less than $\frac
{1}{2}D(D-1)+3$
 angular momenta.

There are $3nj$-symbols of the first and second kind, depending on the
possibility to
bound the first and the last momenta \cite{JB}(see fig.1). We shall consider
$3nj$-symbols
of only the first kind. The application of $3nj$-symbols of the second kind is
the same.

Any $3nj$-symbol can be represented as a sum of products of the corresponding
$6j$-symbols.
\begin {eqnarray}
\left\{ \begin{array}{cccc} j_1 & j_2 & ... & j_n \\ l_1 & l_2 & ... & l_n \\
k_1 & k_2 & ... &
k_n\end{array}\right\} =\sum _x(-1)^{R_n+(n-1)x}(2x+1)
\left\{\begin{array}{ccc} j_1 & k_1 & x \\ k_2 & j_2 &
l_1\end{array}\right\}\times
\nonumber\\
\times
\left\{\begin{array}{ccc} j_2 & k_2 & x \\ k_3 & j_3 &
l_2\end{array}\right\}\times...
\times
\left\{\begin{array}{ccc} j_{n-1} & k_{n-1} & x \\ k_n & j_n &
l_{n-1}\end{array}\right\}
\left\{\begin{array}{ccc} j_n & k_n & x \\ j_1 & k_1 & l_n\end{array}\right\},
\end{eqnarray}
where $R_n=\sum^n_{i=1}(j_i+l_i+k_i)$.

Inasmuch as the asymptotic form of the Racah-Wigner $6j$-symbol according to
Ponzano-Regge formula \cite{PR} can be expressed by the euclidean
Einstein-Regge
action $S_R[T]$ for the 3-simplex $T$ (tetrahedron)

\begin{equation}
\left\{\begin{array}{ccc} j_1 & j_2 & j_3 \\ j_4 & j_5 & j_6\end{array}\right\}
\rightarrow \\ exp(i\sum_{m=1}^6j_m \theta _m)=exp(iS_R[T]),
\end{equation}
at $j_m \gg 1$, expanding the $3nj$-symbol into the sum of products of
6j-symbols we
would obtain a correlation between the euclidean action of gravity $S_R$ and
the
$3nj$-symbol.

There are, however, some difficulties in the implementation of this project. In
particular,
since expression (2) is valid only at $j_i
\gg 1$ expanding the $3nj$-symbol into 6j-simbols we are summing by the
variable
moment $x$, which obeys the condition of triangle $|j_i-j_j|
\leq x\leq j_i-j_j$, from which it follows that even at $j_i,j_j \gg 1$,
we may receive $|j_i-j_j|\simeq 1$, and expression (2) becomes invalid.
Consequently, for
any meaning of the space-time dimension $D$ we must choose a special type of
the $3nj$
symbol, at the condition
\begin{equation}
3n > \frac{1}{2}D(D-1)+3,\quad  n >2, \; D>3.
\end{equation}

To escape summing when expanding (1), let us e|ual one of the momenta to zero:
\begin {eqnarray}
\left\{ \begin{array}{cccc} j_1 & j_2 & ... & 0 \\ l_1 & l_2 & ... & l_n \\
k_1 & k_2 & ... &
k_n\end{array}\right\} =\delta (l_n,k_1)\delta (j_{n-1},l_{n-1}) \times
\nonumber\\
\times[(2k_1+1)(2j_{n-1}+1)]
^{-\frac{1}{2}}
 (-1)^{R_{n-1}+nk_n-j_1+k_{n-1}+j_{n-1}} \times \\
\left\{\begin{array}{ccc} j_1 & k_1 & x \\ k_2 & j_2 & l_1\end{array}\right\}
\left\{\begin{array}{ccc} j_2 & k_2 & x\\ k_3 & j_3 & l_2\end{array}\right\}
...\left\{\begin{array}{ccc} j_{n-1} & k_{n-1} & x \\ k_n & j_n &
l_{n-1}\end{array}\right\}\nonumber
\end{eqnarray}

In this case the $3nj$-symbol depends on only $3(n-1)$ momenta.

When the number of edges of the $D$-simplex belonging to subspace $R^3$ is a
number
multiple to 3, the quantity $3n$ in the $3nj$-symbol is
just equal to the number of edges in $R^3$ plus 3, i.e. $3n=\frac
{1}{2}D(D-1)+6$. When
the number of edges in $R^3$ is the number not multiple to 3, then we put
$3n=\frac{1}{2}D(D-1)+8$ and impose additional conditions like $j_i=j_k$ or/and
$j_i=j_{n-1}$ to decrease the number of independent momenta. Some information
about
D-simplices and corresponding $3nj$-symbols for $D\leq 11$ is given in Table 1.

Now, for the reduced $3nj$-symbol we use equation (1) and the expression of
Ponzano-Regge (2):
\begin{eqnarray}
\left\{\begin{array}{c}reduced \\ 3nj \end{array}\right\} \mapsto
\frac{N}{(12\pi )^{\frac{n-2}{2}}\prod ^{\frac{n-2}{2}}_{k=1}V_k^{\frac{1}{2}}
} \times
\nonumber \\
\times exp(i\sum^{n-2}_{m=1}j_{m}\theta_m+
i\sum^{n}_{m'=1}j_{m'}\theta_m'+
i\sum^{n}_{m''=2}k_{m''}\theta_m''),
\end{eqnarray}
where $N\equiv N_1=(-1)^{R_{n-1}+nk_n-j_1+k_{n-1}+j_{n-1}}\delta
(l_n,k_1)\delta(j_{n-1},l_{n-1}) \times \\
\times[(2k_1+1)(2j_{n-1}+1)]^{-\frac{1}{2}}$, for
$D\neq 3n+2$ and $N\equiv N_1\delta(j_i,j_{n-1})\delta(j_{i-1},j_{n-2}),\\
i\neq n$ for
$D=3n+2$, $V_k$ - volume of the $k$-th tetrahedron.

Let us introduce new denotations
\begin{eqnarray}
(j_1,...,j_{n-1},l_1,...,l_n,k_2,...,k_n)\rightarrow (J_1,...,J_M), \\
(\theta _m,\theta_{m'},\theta_{m''}) \rightarrow (\psi _1,..., \psi_M),
\end{eqnarray}

Then the reduced $3nj$-symbol is
\begin{eqnarray}
\left\{\begin{array}{c}reduced \\ 3nj \end{array}\right\} \mapsto
\frac{N}{(12\pi )^{\frac{n-2}{2}}\prod ^{\frac{n-2}{2}}_{k=1}V_k^{\frac{1}{2}}
}\times
\nonumber \\
\times exp(i\sum^{M}_{m=1}J_{m}\psi_{m}).
\end{eqnarray}

Let us consider the gaussian integral over the real variables $\psi_i$ for the
real non
singular and symmetric $M\times M$ matrix $\Delta $:
\begin{eqnarray}
Z[J_i,\Delta _{ik}]=\int \left\{\begin{array}{c}reduced \\ 3nj
\end{array}\right\}
exp(-\frac{1}{2}\sum_{i,k}^M \psi_i\Delta_{ik}\psi_k)\prod ^M_{j=1}d\psi_j\quad
{}.
\end{eqnarray}

Using the formula for gaussian integration in the  semiclassical limit $J_i\gg
1$ we find the
following results:
\begin{eqnarray}
Z[J_i,\Delta _{ik}]\sim \frac{\hat{N}}{(det\Delta)^{\frac{1}{2}}}
exp(-\sum^M_{i,k=1}J_i(\Delta^{-1})_{ik}J_k),
\end{eqnarray}
where $\hat{N}=(2\pi)^{M-\frac{n-2}{2}}N(6^{-\frac{n-2}{2}}
\prod ^{\frac{n-2}{2}}_{k=1}V^{\frac{1}{2}}_k)^{-1}$.

For the correlation with euclidian quantum gravity we must find the geometrical
interpretation of this gaussian integral. The geometrical meaning of expression
(9) is more
clear if we note that the graphic representation of the $3nj$-symbol in Fig.1
is a
two-dimensional projection of
the $2n$-simplex. Besides, there is a correlation between $D$ and $D+1$
simplices. Every
$(D+1)$-simplex may be built joining two $D$-simplices, by indentification of
two
$(D-1)$-subsimplices and linking the two last free vertices by a new edge in
$(D+1)$
subspace. This means, in particular, that $(D+1)$-simplex at $D>2$ may be
represented
as a
unification of the tetrahedrons with rigid identification edges. This is a
geometrical
analogue
of expanded (7).

Consider
\begin{eqnarray}
\sum^M_{i,k=1}J_i(\Delta^{-1})_{ik}J_k\equiv \sum^M_{i,k=1}\Theta_{ik}A_{ki},
\end{eqnarray}
where $A_{ki}$ is the area of the 2-face of $D$-simplex containing $J_i$ and
$J_k$, and
$\Theta_{ik}$ is a defect angle if the 2-face containing $J_i$ and $J_k$
belongs to interior
of $D$-simplex, or it is an angle between the outer normals of the two
$(D_1)$-simplices
sharing the $(J_iJ_k)$-face, if this 2-face belongs to the boundary of the
$D$-simplex.

For a general $D$-simplicial manifold ${\it M{^{D}}}$ with up to an arbitrary
term depending
on the edge length the euclidean Einstein-Regge action
\begin{equation}
S_R[M^D]\sim \sum_{\sigma \in int M^D}A(\sigma)\epsilon(\sigma)+\sum_{\sigma
\in
\partial M^D}A(\sigma)\alpha (\sigma),
\end{equation}
where $\sigma$ stands for the 2-simplex where the curvature is concentrated and
$A(\sigma)$ is the area of $\sigma $; $\epsilon(\sigma)$ is the defect angle
associated
with the 2-simplex $\sigma $, and $\alpha (\sigma )$ must be interpreted as the
angle
between the outer normals of the two boundary 3-simplices intersecting at
$\sigma$.

Then, according to (10) and (11), we may write
\begin{equation}
Z[J_i,\Delta _{ik}]\sim
\frac{\hat{N}}{(det\Delta)^{\frac{1}{2}}}
exp(-S_R[\sigma ^D ]),
\end{equation}
where $S_r[\sigma^D]$ is the Einstein-Regge action for the $D$-simplex
$\sigma$. We
may consider this expression as a semiclassical limit of the partition function
for $D$
gravity and as a direct generalization of the Ponzano and Regge and
Carfora-Martellini-Marzuoli result.

In principle, there is another way to generalize the Ponzano-Regge ideas. We
may try to
build $3nj$-symbols for every unitary algebra $su(N)$ and, possible, for their
products and
use the correlation between the action of quantum gravity and a corresponding
$3nj$-symbol.

We could receive useful results only for some particular cases in $D=4$ and
$D=6$ by
using the isomorphisms $so(4)\simeq su(2)\oplus  su(2)$ and $so(6)\simeq
su(4)$. Then,
for $su(4)$ algebra, for instance, $21j$-symbol corresponds to $D=6$-simplex.
In general,
there are several variants for building the euclidean quantum action for higher
$D$ as a
unification of 3 or 4, or ... $D-1$ simplices and corresponding $3nj$- momenta.
For other
dimensions, we must again use $3nj$-symbols at $n>2$. On the other hand, on
this way
we have one interesting possibility to build a pseudoeuclidean quantum gravity
action in
$D=3$ by introducing $6j$-symbols for $su(1,1)$ group and use isomorphism
$su(1,1)\simeq so(2,1)$. In any case we may consider all these possibilities as
a kind of
selfconsistency: various ways of building the $D$-simplex must give us the same
result.

These ideas may be applied to superspace. By analogy with the rotation case,
the super
$6j$-symbol $(s6j)$ for the superalgebra $osp(1|2)$ is a symmetric
transformation
coefficient that relates two bases in an irreducible representation space. The
irreducible
representations of the superalgebra $osp(1|2)$ have been analysed in
\cite{S,B}. The
$s6j$-symbols for the superalgebra $osp(1|2)$ possess, in addition to the usual
tetrahedral
symmetry, the so-called Regge symmetry in some particular cases\cite{D}. Then
we may
introduce $s3nj$-supersymbols and consider the semiclassical limit of the
euclidean
action in the corresponding superspace for $D$ supergravity.

\vspace{5mm}
\newpage
{\Large \bf{Acknowledgements}}
\vspace{5mm}

Author wants to express his gratitude to the Swedish Institute for Grant
304/01 GH/MLH, which gave him the possibility to enjoy the kind
hospitality of Prof. Antti Niemi, Doc. Staffan Yngve and all members of
the Institute of Theoretical Physics, Uppsala University.

\newpage

\vspace{7 mm}

\begin{tabular}{|l|r|r|r|r|r||r|r|l|} \hline
{$D$}&{3}&{4}&{5}&{6}&{7}&{10}&{11}&{$D$}\\
\hline\hline
{$\sharp V$}&{4}&{5}&{6}&{7}&{8}&{11}&{12}&{$D+1$}\\
\hline
{$\sharp E$} &{6}&{10}&{15}&{21}&{28}&{55}&{66}&{$1/2D(D+1)$}\\
\hline
{$\sharp E\in R^3$}&{6}&{9}&{13}&{18}&{24}&{48}&{58}&{$1/2D(D-1)+3$}\\
\hline
{3n(3nj)}&{6}&{12}&{18}&{21}&{27}&{51}&{63}&{$\frac
{1/2D(D-1)+8,\quad D=3n+2}{1/2D(D-1)+6,\quad D-others}$}\\
\hline
{$\sharp $add.cond.}&{-}&{2}&{4}&{2}&{2}&{2}&{4}&{$\frac
{4,\quad D=3n+2}{2,\quad D-others}$}\\
\hline
\end{tabular}

\vspace{3 mm}

{\bf Table 1.} {\small {\bf Some properties of $D$-simplices and
$3nj$-symbols}}.\\
{\footnotesize $D$ is a space-time dimension, $\sharp V$ is the number of
vertices,
$\sharp E$ is the number of edges, $\sharp E \in R^3$ is the number of edges
belonging to
subspace $R^3$, $3n(3nj)$ is the quantity of the $3nj$-symbol, $\sharp
add.cond.$ is the
number of additional conditions like $j_i=j_k$, which we impose on the
corresponding
$3nj$-symbol. One general additional condition $j_n=0$ is valid for all
$3nj$-symbols to
escape summation.}

\vspace{5 mm}

\end{document}